\documentclass[]{zakoproc}
\usepackage{graphicx}
\usepackage{amsmath}
\usepackage{natbib}

\newcommand{\satlev}{\left(\Em/\Ek\right)_\mathrm{sat}}
\newcommand{\solratio}{E_\mathrm{sol}/E_\mathrm{tot}}
\newcommand{\ted}{t_\mathrm{ed}}
\newcommand{\Em}{E_{\mathrm{m}}}

\newcommand{\Ek}{E_{\mathrm{k}}}

\begin{document}

\title{Magnetic fields in the first galaxies: Dynamo amplification and limits from reionization}
\author{Dominik R.G. Schleicher\footnote{Institut f\"ur Astrophysik, Georg-August-Universit\"at G\"ottingen, Friedrich-Hund-Platz 1, 37077 G\"ottingen, Germany}, \quad Jennifer Schober\footnote{Zentrum f\"ur Astronomie der Universit\"at Heidelberg, Institut f\"ur Astrophysik, Albert-Ueberle-Str. 2, 69120 Heidelberg, Germany }, \quad Christoph Federrath$^\dag$\footnote{Ecole Normale Sup\'erieure de
Lyon, F-69364 Lyon, France}, \\ Francesco Miniati\footnote{Physics Department, Wolfgang-Pauli-Strasse 27, ETH-Z\"urich, CH-8093 Z\"urich, Switzerland}, \quad Robi Banerjee\footnote{Hamburger Sternwarte, D-21029 Hamburg, Germany}, \quad Ralf S. Klessen$^\dag$}
\institute{}

\markboth{Schleicher et al.}{Magnetic fields in the first galaxies}

\maketitle\vspace{-0.7cm}

\begin{abstract}
We discuss the amplification of magnetic fields by the small-scale dynamo, a process that could efficiently produce strong magnetic fields in the first galaxies. In addition, we derive constraints on the primordial field strength from the epoch of reionization.
\end{abstract}

\section{Introduction}

In the local Universe, magnetic fields are observed on virtually all scales~\citep[e.g.][]{Beck96}, and observations confirm the presence of magnetic fields also at high redshift. 
Such evidence has been found through the Faraday Rotation imprint of line-of-sight galaxies on distant QSOs~\citep{Bernet08, Kronberg08}, as well as through the far-infrared - radio 
correlation \citep{Murphy09}. They further exist in the intergalactic medium (IGM). For instance, galaxy clusters exhibit $\mu$G fields~\citep{Clarke01} that require non-negligible 
initial seeds~\citep{Banerjee03,Ryu08,Miniati11}.

Even in cosmic voids, weak magnetic fields seem to exist, as suggested by recent gamma-ray
experiments~\citep[][]{Neronov10,tavecchio10,Taylor11}, but see \citet{2011arXiv1106.5494B}.
Various astrophysical mechanisms have been proposed for the origin of
intergalactic and cosmological magnetic fields~\citep{Miniati11b,bertone06,Gnedin00,Ando10}.  In
particular, the model in~\citet{Miniati11b} makes consistent
predictions for the magnetic fields observed in the cosmic voids, providing a relevant seed for subsequent dynamo amplification.

Primordial models of magnetogenesis~\citep[e.g.][]{Grasso01} provide
an alternative scenario. Magnetic fields from the early universe could be potentially strong, and need to be constrained observationally. This concerns in particular inflationary scenarios \citep{Turner88}, the electroweak phase transition \citep{Baym96},
or the QCD phase transition \citep{Quashnock89, Cheng94,
  Sigl97}. 

In this contribution, we  discuss the amplification of magnetic fields via the small-scale dynamo, which may provide a strong tangled magnetic field already in the first galaxies \citep[e.g.][]{Schleicher10c, Sur10, Federrath11}. We further describe upper limits on the primordial field strength from recent reionization data \citep{Schleicher11}.

\section{The small-scale dynamo for different turbulence models}

\begin{figure}[t]
\centerline{\includegraphics[width=0.5\linewidth]{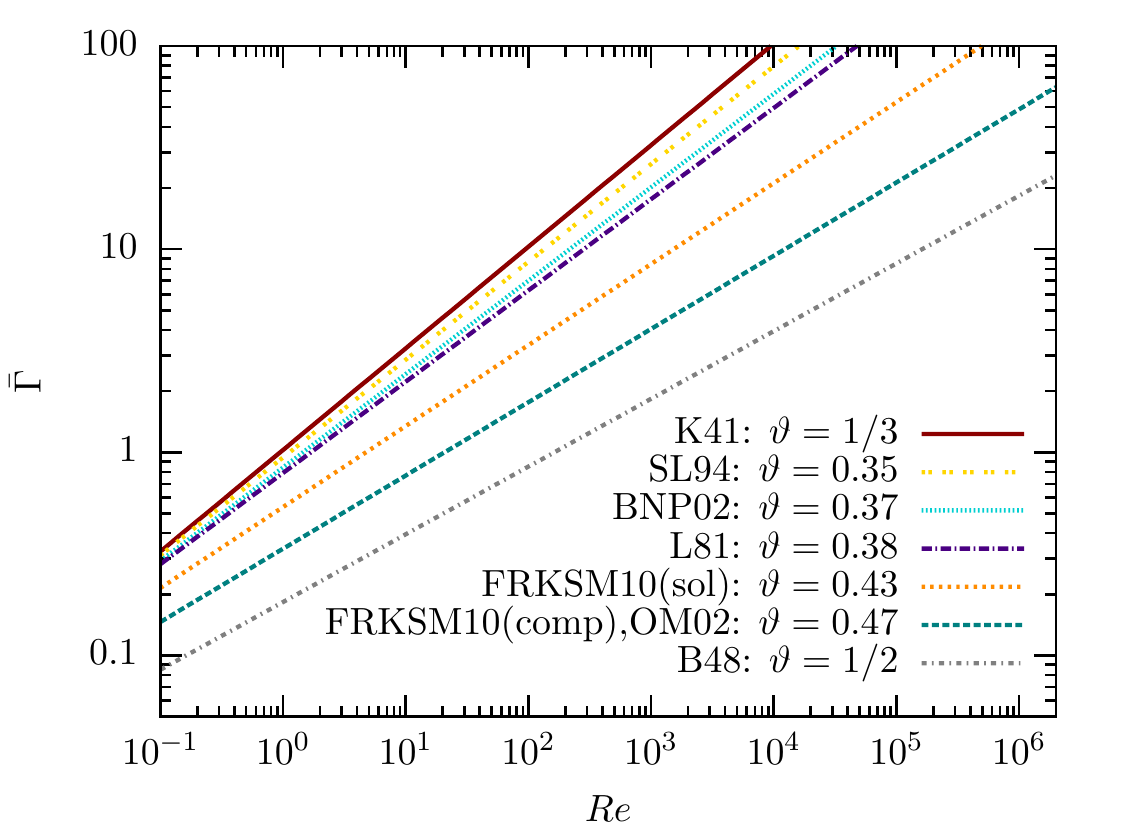}\includegraphics[width=0.5\linewidth]{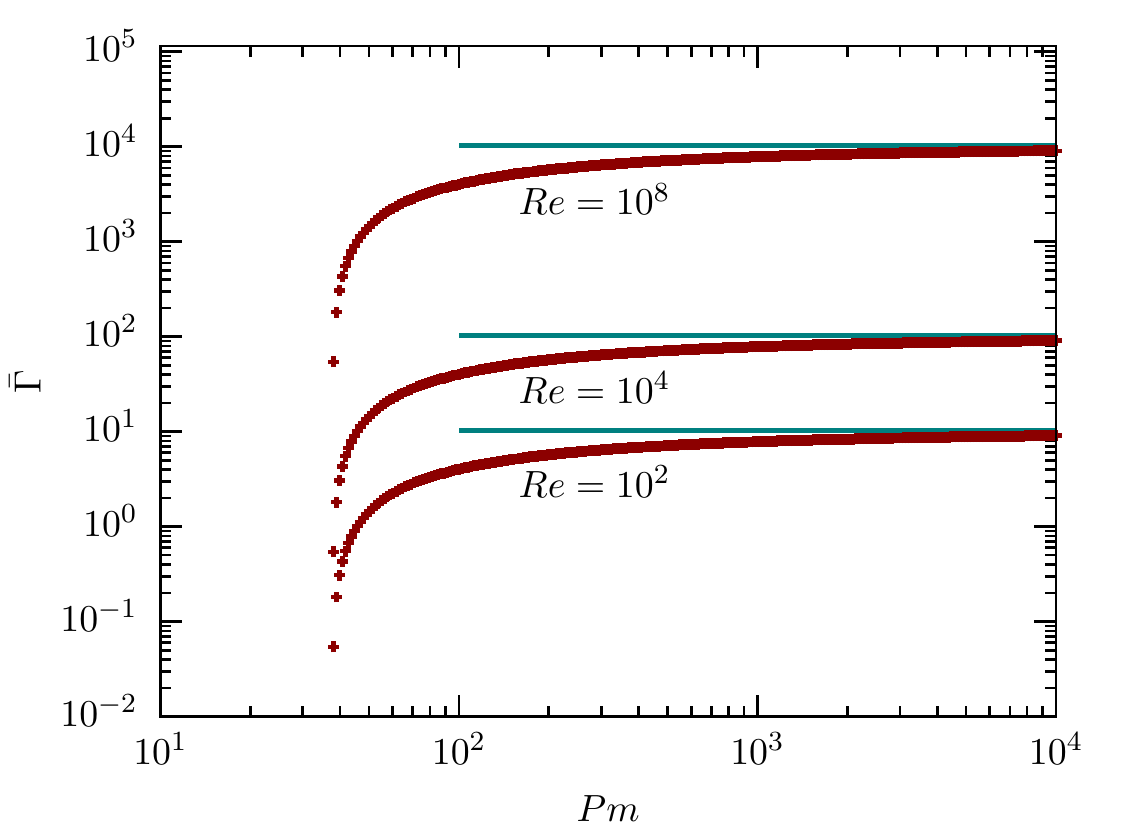}}
\caption{Left: Normalized growth rate for different turbulence models as a function of the Reynolds number, in the limit of an infinte magnetic Prandtl number. Right: Normalized growth rate as a function of the magnetic Prandtl number in the case of incompressible (Kolmogorov) turbulence, for different Reynolds numbers.}
\label{fig:growth}
\end{figure}

The small-scale dynamo provides an efficient amplification mechanism of magnetic fields in the presence of turbulence. It was originally proposed by \citet{Kazantsev68}, and subsequently explored by various authors using analytical models and numerical simulations \citep[e.g.][]{Subramanian97, Subramanian99, Schekochihin02, Haugen04a, Haugen04b, Schekochihin04}. To calculate the growth rate of the magnetic field,
the induction equation, given as
\begin{equation}
  \frac{\partial \textbf{B}}{\partial t} = \nabla\times\textbf{v}\times\textbf{B} - \eta\nabla\times\nabla\times\textbf{B},
\label{induction}
\end{equation}
can be rewritten in terms of the so-called Kazantsev equation 
\begin{equation}
  -\kappa_\text{diff}(r)\frac{\text{d}^2\psi(r)}{\text{d}^2r} + U\psi(r) = -\Gamma \psi(r),
\label{Kazantsev}
\end{equation}
where $\psi$ is related to the spatial dependence of the magnetic field correlation function, $\kappa_\text{diff}$ denotes the turbulent diffusion coefficient and $U$ denotes a function depending on the properties of turbulence. In a recent study by \citet{Schober11}, we solved this equation and derived the dependence of the growth rate on the turbulence model, the magnetic Prandtl number and the Reynolds number. 
For this purpose, we considered the turbulence models given in Table~\ref{ResultsTable}, where we also provide the critical magnetic Reynolds number for magnetic field amplification, as well as the growth rate, which is normalized in terms of the eddy timescale.
In Fig.~\ref{fig:growth}, we show the dependence of the growth rate on the turbulence model, the Prandtl and the Reynolds number. The interested reader is referred to \citet{Schober11} for the derivation of these results.

\begin{table*}  
    \begin{tabular}{l|c|c|c}
      \parbox[0pt][2.5em][c]{0cm}{}  Model/Reference                 & $\vartheta$     &  $Rm_\text{crit}$     & $\bar{\Gamma}$                                                                                                                             ($Pm\rightarrow\infty$)\\
      \hline
      \newline \citet{Kolmogorov41}			    								&  $1/3$          &  $\sim107$     &  $\frac{37}{36}~Re^{1/2}$ \\
      \newline intermittency of Kolmogorov turbulence \citep{SheLeveque94}  &  $0.35$ & $\sim 118$ & $0.94~Re^{0.48}$\\
      \newline driven supersonic MHD-turbulence \citep{Boldyrev02}&  $0.37$  &  $\sim 137$    &  $0.84~Re^{0.46}$     \\
      \newline observation in molecular clouds \citep{Larson81}  &  $0.38$  &  $\sim 149$     &  $0.79~Re^{0.45}$     \\
      \newline solenoidal forcing of the turbulence \citep{Federrath10} &  $0.43$ &  $\sim 227$  &                                                                                                                                  $0.54~Re^{0.40}$  \\
      \newline compressive forcing of the turbulence \citep{Federrath10} &  $0.47$   &  $\sim 697$  &                                                                                                                             $0.34~Re^{0.36}$  \\                        observations in molecular clouds \citep{OssenkopfMacLow02}    &       &       &             \\ 
      \newline \citet{Burgers48}                             &  $1/2$     &  $\sim 2718$   &  $\frac{11}{60}~Re^{1/3}$ \\
\hline  
  \end{tabular}
\caption{The critical magnetic Reynolds number $Rm_\text{crit}$ and the normalised growth rate of the small-scale                                   dynamo $\bar{\Gamma}$ in the limit of infinite magnetic Prandtl numbers. We show our results for different                                 types of turbulence, which are characterised by the exponent $\vartheta$ of the slope of the turbulent                                     velocity spectrum, $v(\ell)\propto\ell^\vartheta$. The extreme values of $\vartheta$ are $1/3$ for Kolmogorov                              turbulence and $1/2$ for Burgers turbulence.}
  \label{ResultsTable}
\end{table*}

\section{The Mach number dependence of magnetic field amplification}
Another quantity with a significant impact on the growth rate and saturation level of the magnetic field is the Mach number of the turbulence. One of the first studies exploring this quantity considered a range of Mach numbers from $0.65$ up to $1.14$ \citep{Haugen04c}.

\begin{figure}[t]
\centerline{\includegraphics[width=0.5\linewidth]{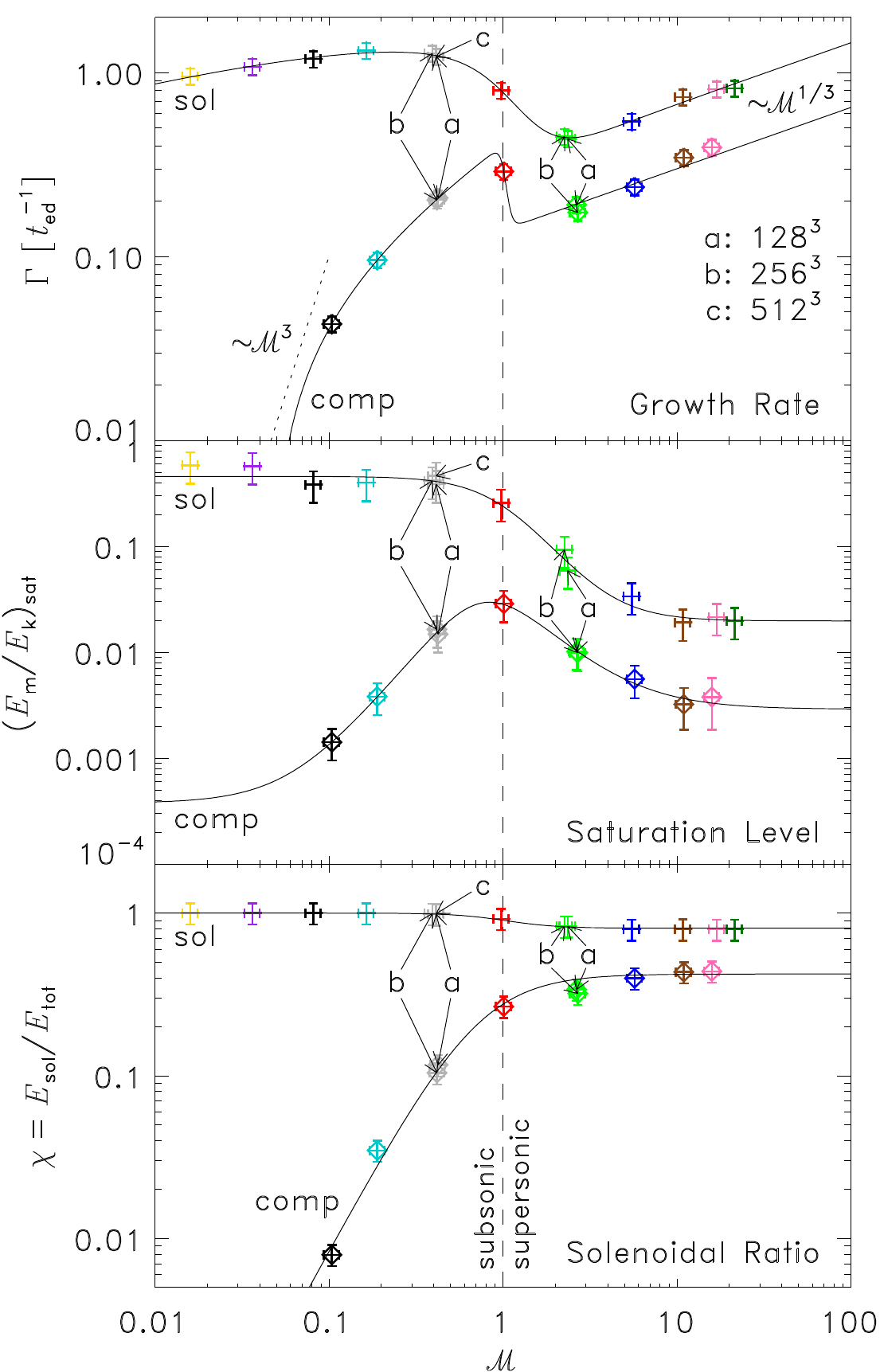}}
\caption{Growth rate (top), saturation level (middle), and solenoidal ratio (bottom) as a function of Mach number, for all runs with solenoidal (crosses) and compressive forcing (diamonds). 
The solid lines show empirical fits given by \citet{FederrathPRL}.}
\label{fig:gratesat}
\end{figure}

We considerably extended the range of Mach numbers in a recent study, exploring the range from Mach $0.02$ up to Mach $20$, using both solenoidal and compressive driving schemes \citep{FederrathPRL}. Our results for the growth rate, the saturation level and the amount of solenoidal turbulent energy are shown in Fig.~\ref{fig:gratesat}.~We derived analytic fits using the fit function
\begin{equation}
\label{eq:fit}
f(x)=\left(p_0\,\frac{x^{p_1}+p_2}{x^{p_3}+p_4}+p_5\right)x^{p_6},
\end{equation}
with the fit parameters given in Table~\ref{tab:fittable}. In the regime of low Mach numbers, all quantities depend considerably on the driving scheme, with solenoidal driving leading to more efficient amplification. Also at high Mach numbers, solenoidal driving is more efficient, but the difference is less pronounced.  For the growth rate of the magnetic field, a transition seems to occur at about Mach 1, where the growth rate drops considerably in the presence of shocks. Towards even larger Mach numbers, we observe an approximate scaling as $\mathcal{M}^{1/3}$ for both types of driving. Overall, our results show that the small-scale dynamo works for a large range of Mach numbers, as well as for compressively driven and solenoidally driven turbulence.

\begin{table}

  \begin{center}
\begin{tabular}{lcccccc}
  & \multicolumn{2}{c}{$\Gamma\,\left[\ted^{-1}\right]$} & \multicolumn{2}{c}{$\satlev$} & \multicolumn{2}{c}{$\solratio$} \\
  & (sol) & (comp) & (sol) & (comp) & (sol) & (comp) \\
\hline
$p_0$ & -18.71                     & \phantom{-}2.251 & \phantom{-}0.020 & \phantom{-}0.037 & \phantom{-}0.808 & \phantom{-}0.423 \\
$p_1$ & \phantom{-}0.051 & \phantom{-}0.119 & \phantom{-}2.340 & \phantom{-}1.982 & \phantom{-}2.850 & \phantom{-}1.970 \\
$p_2$ & -1.059                     & -0.802                     & \phantom{-}23.33 & -0.027                    & \phantom{-}1.238 & 0 \\
$p_3$ & \phantom{-}2.921 & \phantom{-}25.53 & \phantom{-}2.340 & \phantom{-}3.601 & \phantom{-}2.850 & \phantom{-}1.970 \\
$p_4$ & \phantom{-}1.350 & \phantom{-}1.686 & 1                              & \phantom{-}0.395 & 1                              & \phantom{-}0.535 \\
$p_5$ & \phantom{-}0.313 & \phantom{-}0.139 & 0                              & \phantom{-}0.003 & 0                              & 0 \\
$p_6$ & 1/3                           & 1/3                          & 0                              & 0                              & 0                              & 0 \\
\hline
\end{tabular}
\caption{Parameters in Eq.~(\ref{eq:fit}) for the fits in Fig.~\ref{fig:gratesat}.}
\label{tab:fittable}
\end{center}
\end{table}

\section{Upper limits on the primordial field strength}

\begin{figure}[t]
\centerline{\includegraphics[width=0.55\linewidth]{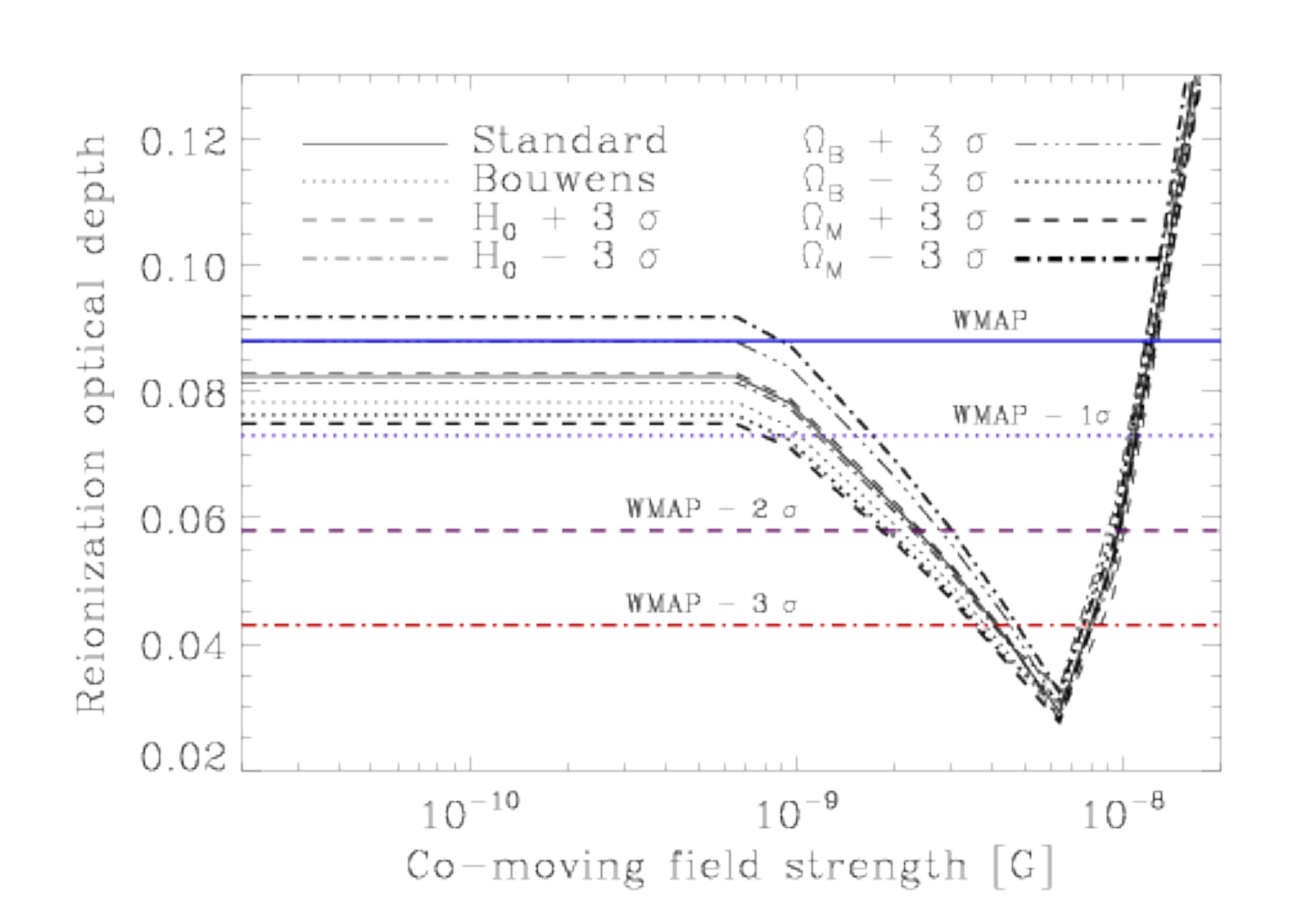}\includegraphics[width=0.55\linewidth]{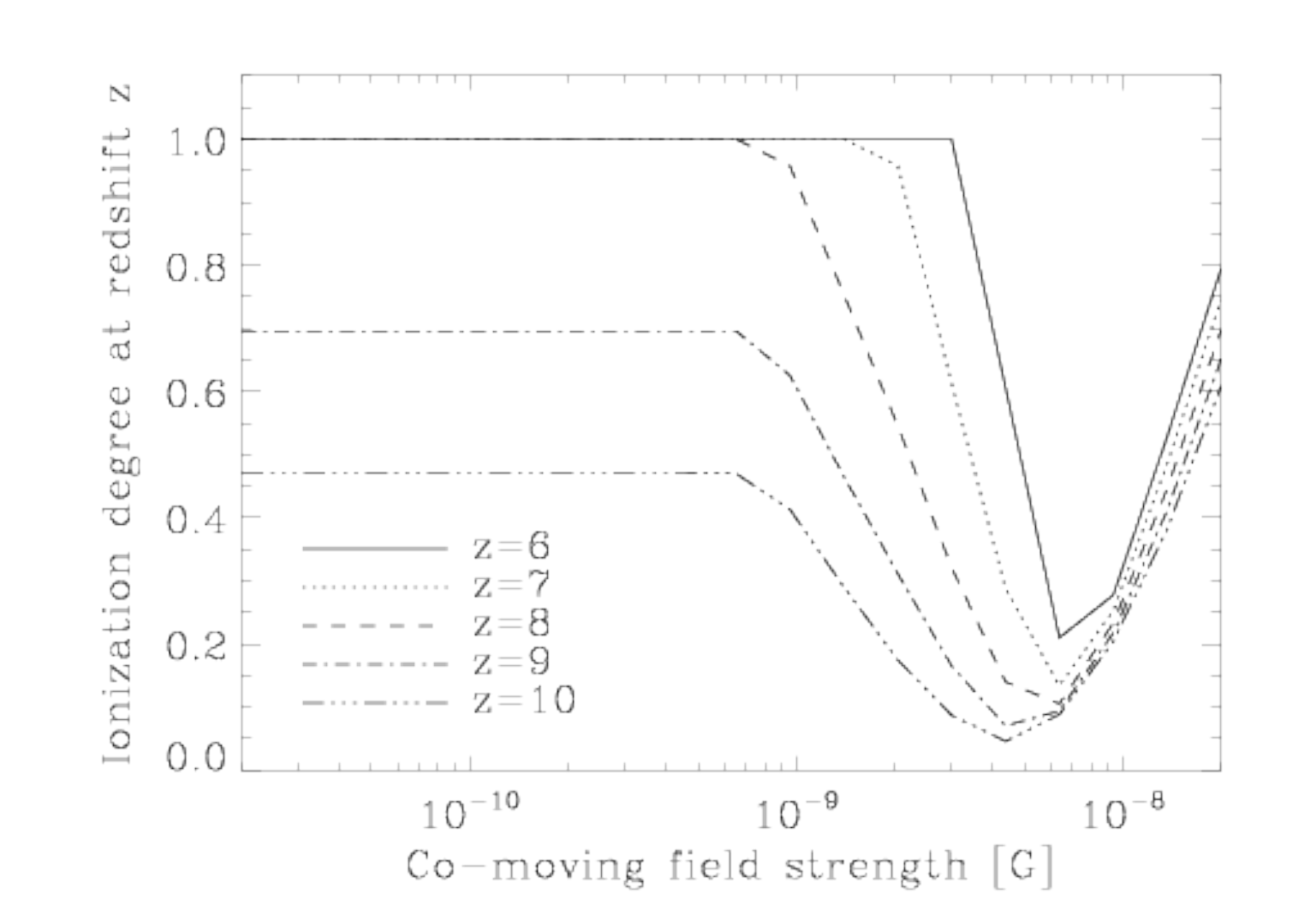}}
\caption{Left: The expected reionization optical depth as a function of the co-moving field strength for different cosmological parameters. Right: The expected ionization degree at different redshifts as a function of the co-moving field strength.}
\label{fig:reion}
\end{figure}

If strong primordial fields have been created in the early Universe, they will subsequently affect structure formation via the magnetic Jeans mass, which is given as \citep{Subramanian98p, Schleicher09prim}
\begin{equation}
M_J^B=10^{10} M_\odot \left(\frac{B_0}{3~nG} \right)^3,\label{JeansB}
\end{equation}
with $B_0$ the co-moving field strength. The latter is related to the physical field strength $B$ via $B_0=B/(1+z)^2$. To explore the implications for the epoch of reionization, we follow the time evolution of the ionized volume fraction $Q_{HII}$ as \citet{Madau99}, yielding
\begin{equation} 
\frac{dQ_{HII}}{dt}=-\frac{Q_{HII}}{t_{rec}}+\frac{{\mathrm SFR}(z)f_{esc}10^{53.2}}{n_H(0)},\label{reion}
\end{equation}
with $t_{rec}$ the recombination timescale, $f_{esc}$ the escape fraction, $n_H(0)$ the co-moving number density. The star formation rate is calculated from the observed Schechter function provided by \citet{Bouwens11}. The further details of this approach are given by \citet{Schleicher11}. From the ionized volume fraction as well as the ionized fraction in the neutral component, we calculate the effective ionization degree $x_{eff}$ and the reionization optical depth
\begin{equation}
\tau_e=\frac{n_H(0) c}{H_0} \int_{z=0}^{z=z_s}x_{eff}(z)\sigma_T \frac{(1+z)^2}{\sqrt{\Omega_\Lambda+\Omega_m(1+z)^3}}dz,
\end{equation}
with $c$ the speed of light, $H_0$ the Hubble constant, $\sigma_T$ the Thompson scattering optical depth and the cosmological density parameters $\Omega_\Lambda$ and $\Omega_m$. We explore the uncertainties due to the cosmological parameters, the reionization parameters as well as the uncertainty in the observed Schechter function on the reionization optical depth, and the ionization degree at different redshifts.
Some of our results are given in Fig.~\ref{fig:reion}, and additional details are provided by \citet{Schleicher11}. Overall, they lead to a $2\sigma$ constraint of $\sim2$~nG on the co-moving field strength.

\acknowledgements{ C.F.~acknowledges funding from a Discovery Projects Fellowship of the Australian Research Council (grant~DP110102191) and from the European Research Council (FP7/2007-2013 Grant Agreement no.~247060). C.F.,~R.B.,~and R.S.K.~acknowledge subsidies from the Baden-W\"urttemberg-Stiftung under research contract P-LS-SPII/18 and from the German Bundesministerium f\"ur Bildung und Forschung via the ASTRONET project STAR FORMAT (grant 05A09VHA). R.S.K also thanks the {\em Deutsche Forschungsgemeinschaft} (DFG) for financial support via grants KL1358/10 and KL1358/11, as well as via the SFB 881 {\em The Milky Way System}. R.B. acknowledges funding by the Emmy-Noether grant (DFG) BA~3706. 
}
\setlength\bibsep{0pt}

\end{document}